\documentclass{article}

\usepackage[preprint]{neurips_2025}

\usepackage[utf8]{inputenc}
\usepackage[T1]{fontenc}
\usepackage{hyperref}
\usepackage{url}
\usepackage{booktabs}
\usepackage{amsfonts}
\usepackage{amsmath}
\usepackage{nicefrac}
\usepackage{microtype}
\usepackage{xcolor}

\title{Domain-Specific Text Embedding Models \\ for Entity Resolution}

\author{%
  S Khajesh Narayana, R Srivardhani, Kishore Reddy Konda \\
  Sodhana \\
\texttt{khajeshn@gmail.com}\\
\texttt{srivardhani.r@nexusiqsolutions.com}\\
\texttt{kishore.konda@sodhana.ai}
}

\begin{document}

\maketitle

\begin{abstract}
General-purpose text embedding models are designed to capture semantic similarity but are not optimised for distinguishing entity records that represent the same real-world business or person. This limitation affects applications such as entity resolution and duplicate record retrieval, where small textual differences may either preserve or change identity. This paper investigates whether domain-specific triplet fine-tuning can adapt pretrained embedding models for identity-sensitive retrieval. A synthetic dataset of business and person records was created with identity-preserving variations and challenging non-matching examples. Two widely used embedding models were evaluated before and after fine-tuning using a margin-based similarity evaluation. The results show substantial improvements in separating true matches from highly similar non-matches, demonstrating that domain-specific triplet training can effectively reshape general-purpose embedding spaces for entity retrieval. These findings suggest that targeted fine-tuning provides a practical approach for improving embedding models in data quality management and information retrieval applications.
\end{abstract}


\section{Introduction}
\label{sec:intro}

Information retrieval systems increasingly rely on dense text embeddings to retrieve semantically similar documents, entities, and records \citep{reimers2019,karpukhin2020}. Modern embedding models have significantly improved retrieval quality by representing text in continuous vector spaces where semantically related inputs are placed close together. While these representations perform well for general semantic search, they are not explicitly designed for applications where retrieval depends on recognizing the identity of real-world entities rather than broad semantic similarity.

Entity resolution is one such application. Organizations routinely maintain large collections of records describing businesses, customers, suppliers, products, and individuals. These records often originate from multiple operational systems, external data sources, manual data entry, or historical databases. As a result, the same real-world entity may appear in several textual forms because of abbreviations, spelling variations, formatting differences, incomplete addresses, punctuation, transliteration, or inconsistent naming conventions. At the same time, different entities may share substantial textual overlap, making reliable retrieval considerably more challenging than conventional semantic search \citep{christen2012}.

Traditional entity resolution approaches commonly rely on rule-based standardization, exact matching, edit distance, token similarity, phonetic algorithms, or manually engineered field-level comparisons \citep{christen2012,elmagarmid2007}. These methods remain valuable because they are interpretable and can effectively capture known textual variations. However, they typically require extensive domain expertise, language-specific rules, and carefully tuned thresholds. Their performance may deteriorate when multiple textual variations occur simultaneously or when highly similar records differ only in a small but decisive component such as a business location, branch identifier, or residential address.

Recent advances in sentence embedding models \citep{reimers2019,xiao2023} offer an attractive alternative by learning dense vector representations that capture contextual relationships without requiring manually defined matching rules. Models such as sentence-transformers/all-MiniLM-L6-v2 and BAAI/bge-base-en-v1.5 have become widely adopted for semantic search and dense retrieval because they generalize well across a broad range of textual domains. However, these models are primarily optimised for semantic relatedness rather than entity identity. Consequently, records referring to different businesses or individuals may receive high similarity scores because they share names, locations, or other contextual information, while genuine duplicate records containing abbreviations or formatting differences may not be sufficiently distinguished from difficult non-matching candidates.

This limitation suggests that entity resolution should be viewed as a specialized representation learning problem. Rather than learning general semantic similarity, embedding models must learn to preserve variations that do not alter identity while simultaneously increasing sensitivity to textual changes that indicate different entities. Achieving this balance is particularly important in large-scale retrieval systems, where small differences in similarity scores can substantially influence candidate ranking and downstream matching decisions.

This paper investigates whether general-purpose sentence embedding models can be adapted for identity-sensitive information retrieval through domain-specific triplet fine-tuning. A synthetic dataset of business and person records was constructed in which positive examples represent realistic identity-preserving variations, and negative examples introduce subtle but meaningful identity or location conflicts. Two widely used embedding models were evaluated before and after fine-tuning using a margin-based similarity evaluation designed to measure how effectively the learned representations separate valid duplicate records from highly similar non-matching entities.

The principal contributions of this work are:
\begin{itemize}
  \item A domain-specific triplet construction methodology that captures realistic identity-preserving and identity-conflicting variations in business and person records.
  \item A margin-based evaluation protocol for measuring identity-sensitive separation in dense embedding spaces.
  \item An empirical study of domain adaptation on two widely used sentence embedding models.
  \item Experimental evidence demonstrating that carefully designed triplet fine-tuning substantially improves the retrieval of duplicate entity records while increasing separation from challenging non-matching candidates.
\end{itemize}

The findings demonstrate that domain-specific supervision can reshape general-purpose embedding spaces into representations that are better suited for entity resolution and duplicate record retrieval. More broadly, the study illustrates how targeted representation learning can bridge the gap between semantic similarity and identity-sensitive retrieval in enterprise information retrieval systems.

\section{Dataset and experimental design}
\label{sec:dataset}

\subsection{Task formulation and triplet structure}

Entity resolution was formulated as a similarity-ranking task over textual representations of business and person records. Given a reference record, the objective is to rank records describing the same real-world entity ahead of highly similar records referring to different entities. This formulation reflects the requirements of dense information retrieval systems, where candidate records are retrieved and ranked according to embedding similarity.

A triplet-based learning framework \citep{schroff2015,hoffer2015} was adopted because it directly optimizes the relative ordering of matching and non-matching records. Rather than learning whether two records are simply similar, the model learns to position identity-preserving variations closer to a reference record than plausible alternatives representing different entities. This objective is naturally suited to nearest-neighbor retrieval using dense vector embeddings.

Each training example consisted of three records:
\begin{itemize}
  \item \textbf{Anchor}: the reference business or person record.
  \item \textbf{Positive}: a variation of the anchor representing the same real-world entity.
  \item \textbf{Negative}: a different entity that remains textually similar to the anchor.
\end{itemize}

Positive examples preserved entity identity while introducing realistic variations commonly encountered in enterprise data, including abbreviations, formatting changes, punctuation differences, reordered address components, and omission of non-essential information. Negative examples were designed to remain plausible and textually similar while introducing meaningful identity or location conflicts.

The dataset primarily emphasized hard and moderately difficult negatives rather than completely unrelated records. Many negative examples intentionally shared company names, personal names, surnames, buildings, cities, or address components with the anchor, requiring the model to distinguish subtle identity differences despite substantial textual overlap. Completely unrelated negatives were avoided because they provide limited supervision for learning fine-grained identity-sensitive representations.

\subsection{Business and person record construction}

To evaluate domain-specific representation learning under controlled conditions, a synthetic dataset of business and person records was constructed. Synthetic data was chosen to provide complete control over entity identity, textual variation, and negative example generation while avoiding the use of confidential or personally identifiable information. The records were designed to resemble realistic international naming conventions and address formats commonly encountered in enterprise information systems.

Two categories of entity records were generated. Business records combined fictional company names with structured commercial addresses, while person records combined fictional personal names with structured residential addresses. Both record types followed a natural language template to produce complete textual descriptions suitable for input to sentence embedding models:

\begin{quote}
Business records: \{Company Name\} is a firm located at \{Address\}

Person records: \{Person Name\} is a person living at \{Address\}
\end{quote}

Addresses were constructed from combinations of unit identifiers, building names or numbers, street names, cities, regions, postal codes, and countries. Record generation incorporated variations in legal entity designations, address abbreviations, punctuation, word order, and formatting to reflect the diversity typically observed in real-world enterprise data.

The dataset consisted of independent training and validation splits, with entity anchors separated before triplet generation to ensure that validation entities were not observed during training. Table~\ref{tab:dataset} summarises the number of business and person anchors together with the corresponding triplets generated for each dataset partition.

\begin{table}[t]
  \caption{Dataset summary.}
  \label{tab:dataset}
  \centering
  \begin{tabular}{lccccc}
    \toprule
    Split & Business Anchors & Person Anchors & Business Triplets & Person Triplets & Total Triplets \\
    \midrule
    Training   & 313 & 300 & 3{,}130 & 3{,}000 & 6{,}130 \\
    Validation & 100 & 100 & 1{,}000 & 1{,}000 & 2{,}000 \\
    Overall    & 413 & 400 & 4{,}130 & 4{,}000 & 8{,}130 \\
    \bottomrule
  \end{tabular}
\end{table}

Table~\ref{tab:dataset} reports the number of unique anchors and generated triplets in each split. Training and validation anchors were separated before triplet generation, so the 2,000 validation triplets were derived from entities not used to create the training set.

\subsection{Positive and negative example construction}

The effectiveness of triplet learning depends on the quality of the positive and negative examples. Rather than applying arbitrary text augmentation, each triplet was designed to reflect realistic variations encountered in entity records while preserving clear supervision regarding entity identity.

Positive examples represented the same real-world business or person as the anchor while introducing variations commonly observed in enterprise data. These variations included abbreviations of legal entity designations and address components, punctuation and capitalization changes, reordered address elements, omission of non-essential fields such as postal codes or countries, and variations in personal names, including initials, transliterations, and diacritic normalization. Each positive example was reviewed to ensure that the underlying entity identity and intended location remained unchanged.

Negative examples represented different entities while remaining intentionally similar to the anchor. To create challenging retrieval scenarios, negatives frequently shared company names, personal names, surnames, buildings, cities, or address components with the corresponding anchor. Identity conflicts were introduced through changes such as different business locations, branch-level distinctions, different residential addresses, or different individuals with similar names. This design encourages the model to learn fine-grained identity distinctions rather than relying on broad semantic similarity.

The following examples illustrate the triplet construction process.

\textbf{Business example}
\begin{itemize}
  \item Anchor: Horizon Technologies Limited is a firm located at Suite 4, 28 King Street, London, United Kingdom.
  \item Positive: Horizon Technologies Ltd is a firm located at Ste 4, 28 King St, London, UK.
  \item Negative: Horizon Technologies Limited is a firm located at 86 Queen Street, Manchester, United Kingdom.
\end{itemize}

The positive preserves the same company and location despite changes in legal entity designation, address abbreviations, and formatting, whereas the negative retains the company name but represents a different location.

\textbf{Person example}
\begin{itemize}
  \item Anchor: Elena Okonkwo is a person living at Apartment 12, 45 Oak Road, Lagos, Nigeria.
  \item Positive: E.\ Okonkwo is a person living at Apt 12, 45 Oak Rd, Lagos, Nigeria.
  \item Negative: Elena Okonkwo is a person living at Flat 7, 91 Palm Street, Lagos, Nigeria.
\end{itemize}

The positive represents the same individual using abbreviated personal and address information, whereas the negative preserves the person's name while introducing a different residential location.

To maximize the training signal, the dataset primarily employed hard and moderately difficult negatives. Records differing only in punctuation, capitalization, or minor formatting variations were not treated as negatives unless accompanied by a genuine identity or location conflict. This reduces the likelihood of the model learning superficial textual cues and instead encourages the development of identity-sensitive embedding representations.

\subsection{Data separation and quality assurance}

To minimize the risk of memorization and information leakage, training and validation data were separated at the entity level before triplet generation. Business and person entities used to construct the validation set were excluded entirely from the training dataset, ensuring that evaluation measured the models' ability to generalize to previously unseen entities rather than recall memorized examples.

Following dataset generation, both automated validation and manual review were performed to verify data quality. Duplicate records, invalid triplets, inconsistent labels, and unrealistic transformations were identified and corrected. Positive examples were checked to ensure that they preserved the intended entity identity and location, while negative examples were verified to contain meaningful identity or location conflicts.

Additional quality assurance focused on reducing shortcut learning. The dataset was examined for repeated transformation patterns, structural artefacts, and class imbalances that could allow models to distinguish positives from negatives using superficial textual cues instead of learning identity-sensitive representations. These validation procedures were intended to ensure that the reported results reflect meaningful representation learning rather than dataset bias or information leakage.

\subsection{Embedding models}

Two widely used pretrained sentence embedding models were selected to evaluate whether domain-specific triplet learning consistently improves identity-sensitive retrieval across different model architectures. The objective was not to identify the best general-purpose embedding model, but to investigate whether the proposed training methodology could effectively adapt existing semantic embedding models to the entity resolution task.

The models evaluated were:
\begin{itemize}
  \item sentence-transformers/all-MiniLM-L6-v2, developed within the Sentence-Transformers framework \citep{reimers2019,minilm};
  \item BAAI/bge-base-en-v1.5 \citep{xiao2023,bge}.
\end{itemize}

The exact pretrained checkpoints correspond to the publicly available Hugging Face model cards. These models were chosen because they are widely used for semantic similarity and dense information retrieval, providing representative baselines for evaluating the effectiveness of domain adaptation.

Each model was first evaluated using its original pretrained weights to establish a baseline. The models were then independently fine-tuned using the same triplet dataset and training configuration before being evaluated again using the margin-based protocol described in Section~\ref{sec:eval}. This experimental design isolates the effect of domain-specific fine-tuning while allowing each model to be compared against its own pretrained baseline.

\section{Methodology}
\label{sec:method}

\subsection{Evaluation protocol}
\label{sec:eval}

The effectiveness of each embedding model was evaluated using the validation dataset containing 2,000 triplets. Each triplet consisted of an anchor record, an identity-preserving positive example, and a challenging negative example representing a different entity. For each record, dense vector embeddings were generated using the evaluated model and normalized prior to similarity computation. Cosine similarity was then calculated between the anchor and its corresponding positive and negative records:

\begin{equation}
  S_{\text{positive}} = \mathrm{sim}(a, p), \qquad
  S_{\text{negative}} = \mathrm{sim}(a, n),
  \label{eq:sim}
\end{equation}

where $a$, $p$, and $n$ denote the anchor, positive, and negative embeddings respectively, and $\mathrm{sim}(\cdot)$ represents cosine similarity.

For every triplet, the separation between the positive and negative examples was measured as

\begin{equation}
  \text{Score Gap} = S_{\text{positive}} - S_{\text{negative}},
  \label{eq:gap}
\end{equation}

A positive score gap indicates that the matching record was ranked above the non-matching record. However, correct ranking alone does not necessarily imply reliable retrieval performance. When the score difference is small, matching and non-matching records remain close in the embedding space, making candidate retrieval and threshold selection more difficult.

To evaluate the quality of the learned representations, three score-gap margins (0.10, 0.20, and 0.30) were considered. A triplet was regarded as correctly separated when

\begin{equation}
  \text{Score Gap} \geq m,
  \label{eq:margin}
\end{equation}

where $m$ denotes the selected margin. The pass rate was calculated as the percentage of validation triplets satisfying each margin threshold. Larger margins correspond to stronger separation between matching and non-matching records, providing a more demanding measure of identity-sensitive retrieval performance. Table~\ref{tab:margins} summarizes the interpretation of the evaluated margins.

\begin{table}[t]
  \caption{Margin interpretation.}
  \label{tab:margins}
  \centering
  \begin{tabular}{cl}
    \toprule
    Margin & Interpretation \\
    \midrule
    0.10 & Moderate positive--negative separation \\
    0.20 & Strong positive--negative separation \\
    0.30 & Very strong positive--negative separation \\
    \bottomrule
  \end{tabular}
\end{table}

\subsection{Fine-tuning configuration}

Both embedding models were fine-tuned independently using the training dataset described in Section~\ref{sec:dataset}. The objective was to adapt general-purpose semantic representations to distinguish identity-preserving textual variations from challenging non-matching records.

Training employed triplet loss with cosine distance. For an anchor embedding $a$, positive embedding $p$, negative embedding $n$, cosine-distance function $d_{\mathrm{cos}}$, and training margin $m$, the optimization objective was

\begin{equation}
  L = \max\big(d_{\mathrm{cos}}(a, p) - d_{\mathrm{cos}}(a, n) + m,\ 0\big).
  \label{eq:loss}
\end{equation}

The loss becomes positive whenever the positive example is not sufficiently closer to the anchor than the negative example, encouraging the model to reduce the anchor-positive distance while increasing the anchor-negative distance. Optimization therefore directly learns an embedding space in which identity-preserving records are positioned closer together than challenging non-matching entities.

Both models were trained using the same dataset and identical training procedure to ensure a fair comparison. Each model was initialized from its publicly available pretrained weights and fine-tuned independently. Performance was evaluated only on the held-out validation dataset after completion of training.

\section{Results and discussion}
\label{sec:results}

\subsection{Performance before domain adaptation}

Table~\ref{tab:pretrained} presents the performance of the pretrained embedding models on the validation dataset. Both models were generally able to rank identity-preserving records above challenging non-matching records. However, their ability to create clear separation between positive and negative examples was limited, particularly at higher score-gap margins.

At the strictest margin of 0.30, BGE Base EN v1.5 achieved a pass rate of 15.25\%, while all-MiniLM-L6-v2 achieved 37.85\%. Performance improved as the margin requirement became less restrictive, indicating that although the pretrained models frequently assigned higher similarity to matching records, the resulting separation was often small. Consequently, highly similar non-matching entities remained close to genuine duplicates in the embedding space.

These observations are consistent with the objectives of general-purpose sentence embedding models. Such models are trained to capture semantic relatedness across diverse domains rather than distinguish fine-grained entity identity. Records sharing company names, personal names, or address vocabulary therefore tend to receive similar representations even when they correspond to different real-world entities.

\begin{table}[t]
  \caption{Validation performance of pretrained models across score-gap margins.}
  \label{tab:pretrained}
  \centering
  \begin{tabular}{lccc}
    \toprule
    Model & Margin 0.30 & Margin 0.20 & Margin 0.10 \\
    \midrule
    BGE Base EN v1.5 (pretrained)  & 15.25\% & 41.85\% & 60.85\% \\
    all-MiniLM-L6-v2 (pretrained) & 37.85\% & 49.35\% & 64.25\% \\
    \bottomrule
  \end{tabular}
\end{table}

\subsection{Effect of domain-specific fine-tuning}

Table~\ref{tab:finetuned} compares the pretrained and fine-tuned models under the same evaluation protocol. Fine-tuning produced substantial improvements across all evaluated margins for both architectures, demonstrating that domain-specific supervision successfully adapted the embedding space for identity-sensitive retrieval.

The largest improvements were observed at the most demanding margin of 0.30. BGE Base EN v1.5 improved from 15.25\% to 92.70\%, while all-MiniLM-L6-v2 improved from 37.85\% to 83.10\%. Similar improvements were observed at margins of 0.20 and 0.10, where both models exceeded their respective pretrained baselines by a considerable margin.

These results indicate that triplet fine-tuning did more than preserve the ranking of positive examples. The learned representations increased the distance between identity-preserving records and challenging non-matching records, producing substantially clearer separation within the embedding space. Such separation is particularly valuable for nearest-neighbor retrieval, where small differences in similarity scores can strongly influence candidate ranking.

\begin{table}[t]
  \caption{Pretrained and fine-tuned validation performance.}
  \label{tab:finetuned}
  \centering
  \begin{tabular}{lccc}
    \toprule
    Model & Margin 0.30 & Margin 0.20 & Margin 0.10 \\
    \midrule
    BGE Base EN v1.5 (pretrained) & 15.25\% & 41.85\% & 60.85\% \\
    BGE Base EN v1.5 (fine-tuned) & 92.70\% & 96.85\% & 98.60\% \\
    all-MiniLM-L6-v2 (pretrained) & 37.85\% & 49.35\% & 64.25\% \\
    all-MiniLM-L6-v2 (fine-tuned) & 83.10\% & 88.95\% & 93.10\% \\
    \bottomrule
  \end{tabular}
\end{table}

\subsection{Comparative analysis}

Although all-MiniLM-L6-v2 provided the stronger pretrained baseline, BGE Base EN v1.5 exhibited greater improvement following domain-specific fine-tuning. After training, BGE Base EN v1.5 outperformed all-MiniLM-L6-v2 by 9.60 percentage points at margin 0.30, 7.90 percentage points at margin 0.20, and 5.50 percentage points at margin 0.10.

These findings suggest that the two models possess different adaptation characteristics. The stronger pretrained performance of all-MiniLM-L6-v2 indicates better initial generalization for this task, whereas the larger gains achieved by BGE Base EN v1.5 demonstrate a greater capacity to incorporate domain-specific supervision under the proposed training strategy. While these results should not be interpreted as evidence of universal superiority, they indicate that model architecture influences the extent to which domain adaptation improves identity-sensitive retrieval.

\subsection{Practical implications}

The experimental results demonstrate that carefully constructed triplet supervision can substantially improve dense embedding models for entity resolution. Rather than relying solely on general semantic similarity, the fine-tuned models learned representations that tolerate identity-preserving textual variation while increasing sensitivity to meaningful identity and location conflicts.

\subsection{Interpretation of model behavior}

From an operational perspective, the embeddings can support candidate retrieval, ranking of likely duplicates, and prioritization of ambiguous records for manual review. They should be used as one component of a larger deduplication process, together with structured identifiers, postal consistency, business rules, and review policies. The experimental margins are comparison thresholds rather than automatically selected production cut-offs.

\section{Conclusion}
\label{sec:conclusion}

This study investigated the adaptation of general-purpose sentence embedding models for identity-sensitive information retrieval through domain-specific triplet learning. The results demonstrate that embedding spaces can be systematically reshaped to capture application-specific notions of similarity, enabling models to distinguish between textual variations that preserve entity identity and those that represent meaningful identity or location conflicts.

Although the experimental results demonstrate substantial improvements in identity-sensitive retrieval, several limitations should be acknowledged. First, the study relies on a synthetically generated dataset designed to emulate realistic business and person records. While this enables precise control over identity-preserving and identity-conflicting variations, real-world enterprise datasets may exhibit additional sources of noise, ambiguity, and distributional shift. Second, the experimental evaluation is limited to two widely used sentence embedding models, and the observed improvements may not generalize uniformly to other architectures. Finally, the evaluation focuses on embedding-space separation under a controlled retrieval setting rather than complete end-to-end entity resolution pipelines. Future work will therefore investigate larger-scale evaluations using public and industrial entity-resolution benchmarks, additional embedding architectures, and downstream retrieval and matching systems.

Beyond the specific task of entity resolution, these findings suggest a broader perspective on representation learning for information retrieval. Current embedding models typically learn a single semantic representation for a given input, assuming that similarity is fixed across applications. However, in many real-world retrieval problems, the notion of similarity is inherently contextual. Depending on the retrieval objective, the same textual record may require different representations, with certain attributes emphasized, ignored, or treated as decisive.

One possible direction for future research is therefore the development of context-conditioned embedding models, in which the retrieval context forms part of the model input alongside the textual content. Rather than relying solely on fine-tuning to encode a particular notion of similarity, such models could generate embeddings conditioned on explicit contextual instructions describing what aspects of the input should be preserved or distinguished. For example, a retrieval context may specify that legal entity designations and abbreviations should be ignored while address changes remain significant, producing an embedding tailored to that retrieval objective.

This paradigm has the potential to move beyond task-specific embedding models towards more adaptive retrieval systems capable of generating multiple representations of the same text for different information needs. Exploring how contextual guidance can be incorporated into embedding models, and how such representations can be learned efficiently, represents a promising direction for future research in information retrieval and representation learning.

\section*{Code and Data Availability}

The source code, synthetic dataset generation pipeline, trained models, and evaluation scripts used in this study are publicly available at
\url{https://github.com/khajeshnarayana/Domain-Specific-Text-Embedding-Models-for-Information-Retrieval}.

\bibliographystyle{plainnat}

% ============================================================
% APPENDIX
% ============================================================

\appendix
\renewcommand{\thesection}{Appendix \Alph{section}}

% ============================================================
% APPENDIX A
% ============================================================

\section{Training Configuration}
\label{sec:appendix-training}

The same principal settings were used for both models during fine-tuning.
Table~\ref{tab:training-config} summarizes the training configuration.

\begin{table}[h]
  \caption{Training configuration.}
  \label{tab:training-config}
  \centering
  \begin{tabular}{ll}
    \toprule
    Parameter & Value \\
    \midrule
    Training triplets       & 6{,}130 \\
    Validation triplets     & 2{,}000 \\
    Learning rate            & $2 \times 10^{-5}$ \\
    Training batch size      & 16 \\
    Evaluation batch size    & 8 \\
    Number of epochs         & 3 \\
    Loss function            & Triplet loss \\
    Distance metric          & Cosine distance \\
    Triplet margin           & 0.30 \\
    Warm-up ratio            & 0.10 \\
    Evaluation frequency     & Every 200 steps \\
    Checkpoint frequency     & Every 200 steps \\
    Logging frequency        & Every 10 steps \\
    Training seed            & 42 \\
    Data seed                & 42 \\
    Best-model metric        & Triplet cosine accuracy \\
    Early stopping           & Not used \\
    \bottomrule
  \end{tabular}
\end{table}

% ============================================================
% APPENDIX B
% ============================================================

\section{Base vs. Fine-Tuned Model Comparisons}
\label{sec:appendix-comparisons}

This appendix presents representative anchor--positive--negative triplets
used to illustrate the difference between the pretrained and fine-tuned
embedding models. Each example contains an anchor record, a positive
example representing the same entity, and a negative example representing
a near-duplicate distractor. The reported gap is defined as

\begin{equation}
\text{Gap}
=
S_{\text{positive}}
-
S_{\text{negative}}.
\end{equation}

A larger positive gap indicates stronger separation between the true match
and the non-matching candidate.

% ============================================================
% MINILM BUSINESS EXAMPLES
% ============================================================

\subsection{all-MiniLM-L6-v2: Business Entities}

\paragraph{Example 1}

\textbf{Anchor:} Cenquay \& Kin S.A.C. is a firm located at Oficina 172, 92 Plum Fen Jiron, Trujillo, Arequipa 04001, Peru.

\textbf{Positive:} Cenquay \& Kin S.A.C. is a firm located at Of. 172, 92 Plum Fen Jiron, Trujillo, Arequipa 04001, Peru.

\textbf{Negative:} Cenuay \& Kin S.A.C. is a firm located at Oficina 172, 92 Plum Fen Jiron, Trujillo, Arequipa 04001, Peru.

\begin{center}
\small
\begin{tabular}{lccc}
\toprule
Stage & Pos. Similarity & Neg. Similarity & Gap \\
\midrule
Base Model & 0.9820 & 0.8967 & +0.0854 \\
Fine-Tuned & 0.9950 & 0.5280 & +0.4671 \\
\bottomrule
\end{tabular}
\end{center}

\paragraph{Example 2}

\textbf{Anchor:} Society of Xenovara bicycle components Ltd is a firm located at Suite 288, 334 Clover Sena Lane, Bristol, England YO1 7PR, United Kingdom.

\textbf{Positive:} Society of Xenovara bicycle components Ltd is a firm located at Ste 288 - 334 Clover Sena Ln, Bristol, England YO1 7PR, United Kingdom.

\textbf{Negative:} Society of Xenovara bicycle components Ltd is a firm located at Suite 63, Trellis Business Centre, Bristol, England YO1 7PR, United Kingdom.

\begin{center}
\small
\begin{tabular}{lccc}
\toprule
Stage & Pos. Similarity & Neg. Similarity & Gap \\
\midrule
Base Model & 0.9821 & 0.9433 & +0.0389 \\
Fine-Tuned & 0.9962 & 0.5838 & +0.4124 \\
\bottomrule
\end{tabular}
\end{center}

\paragraph{Example 3}

\textbf{Anchor:} Belnori Navigation Guild Ltd. is a firm located at Suite 110, 181 Olive Yara Road, Tamale, Ashanti AK-039-5021, Ghana.

\textbf{Positive:} Belnori Navigation Guild Ltd. is a firm located at Ste 110, 181 Olive Yara Rd, Tamale, Ashanti AK-039-5021, Ghana.

\textbf{Negative:} Belori Navigation Guild Ltd. is a firm located at Suite 110, 181 Olive Yara Road, Tamale, Ashanti AK-039-5021, Ghana.

\begin{center}
\small
\begin{tabular}{lccc}
\toprule
Stage & Pos. Similarity & Neg. Similarity & Gap \\
\midrule
Base Model & 0.9885 & 0.9187 & +0.0698 \\
Fine-Tuned & 0.9968 & 0.5578 & +0.4390 \\
\bottomrule
\end{tabular}
\end{center}

\paragraph{Example 4}

\textbf{Anchor:} Warovara, botanical extracts S.R.L. is a firm located at Apto. 241, 251 Pearl Koa Calle, Salto, Maldonado 20000, Uruguay.

\textbf{Positive:} Warovara, botanical extracts is a firm located at Apto 241 251 Pearl Koa Calle Salto Maldonado 20000 Uruguay.

\textbf{Negative:} Warovara, botanical extracts S.R.L. is a firm located at Apto. 241, 251 Pearl Koa Calle, Salto, Dahlia Region 10012, Uruguay.

\begin{center}
\small
\begin{tabular}{lccc}
\toprule
Stage & Pos. Similarity & Neg. Similarity & Gap \\
\midrule
Base Model & 0.9651 & 0.9822 & -0.0170 \\
Fine-Tuned & 0.8078 & 0.4635 & +0.3443 \\
\bottomrule
\end{tabular}
\end{center}

\paragraph{Example 5}

\textbf{Anchor:} Zurivara \& Kin SARL is a firm located at Bureau 382, 500 Mango Gra rue, Nimes, Occitanie 31000, France.

\textbf{Positive:} Zurivara \& Kin SARL is a firm located at Nimes, Occitanie 31000, Bureau 382, 500 Mango Gra rue, France.

\textbf{Negative:} Zurivara \& Kin SARL is a firm located at Bureau 382, 500 Mango Gra rue, Paris, Occitanie 31000, France.

\begin{center}
\small
\begin{tabular}{lccc}
\toprule
Stage & Pos. Similarity & Neg. Similarity & Gap \\
\midrule
Base Model & 0.9867 & 0.9891 & -0.0025 \\
Fine-Tuned & 0.9952 & 0.6588 & +0.3365 \\
\bottomrule
\end{tabular}
\end{center}

% ============================================================
% MINILM PERSON EXAMPLES
% ============================================================

\subsection{all-MiniLM-L6-v2: Person Entities}

\paragraph{Example 1}

\textbf{Anchor:} Goran Valente is a person living at Unit 55E, 550 Ginger Zel Road, Adama, Amhara 1000, Ethiopia.

\textbf{Positive:} Goran V. is a person living at Unit 55E, 550 Ginger Zel Road, Adama, Amhara 1000.

\textbf{Negative:} Goran Valente is a person living at Unit 123, 435 Jacaranda Glen Lane, Hamilton, Canterbury 8011, New Zealand.

\begin{center}
\small
\begin{tabular}{lccc}
\toprule
Stage & Pos. Similarity & Neg. Similarity & Gap \\
\midrule
Base Model & 0.7644 & 0.7110 & +0.0534 \\
Fine-Tuned & 0.6330 & 0.2020 & +0.4311 \\
\bottomrule
\end{tabular}
\end{center}

\paragraph{Example 2}

\textbf{Anchor:} Nuru Tadesse is a person living at Wohnung 468, 282 Beech Xan Street, Rustavi, Imereti 4600, Georgia.

\textbf{Positive:} Nuru T. is a person living at Wohnung 468, 282 Beech Xan Street, Rustavi, Imereti 4600.

\textbf{Negative:} Nuru Tadesse is a person living at Wohnung 33, 167 Willow Ero Avenue, Salem, Oregon OR 97401, United States.

\begin{center}
\small
\begin{tabular}{lccc}
\toprule
Stage & Pos. Similarity & Neg. Similarity & Gap \\
\midrule
Base Model & 0.7502 & 0.7720 & -0.0218 \\
Fine-Tuned & 0.7974 & 0.4500 & +0.3473 \\
\bottomrule
\end{tabular}
\end{center}

\paragraph{Example 3}

\textbf{Anchor:} Parisa Dervishi is a person living at Apartamento 83, 476 Azalea Lun Road, Mount Hagen, Morobe 411, Papua New Guinea.

\textbf{Positive:} Parisa D. is a person living at Apartamento 83, 476 Azalea Lun Road, Mount Hagen, Morobe 411.

\textbf{Negative:} Parisa Dervishi is a person living at Apartamento 151, 361 Orchard Tovi ulica, Nitra, Zilina 010 01, Slovakia.

\begin{center}
\small
\begin{tabular}{lccc}
\toprule
Stage & Pos. Similarity & Neg. Similarity & Gap \\
\midrule
Base Model & 0.7193 & 0.7732 & -0.0539 \\
Fine-Tuned & 0.7059 & 0.3939 & +0.3120 \\
\bottomrule
\end{tabular}
\end{center}

\paragraph{Example 4}

\textbf{Anchor:} Leire Okafor is a person living at Casa 8, 486 Slate Bri Avenida, Heredia, San Jose 10101, Costa Rica.

\textbf{Positive:} Leire O. is a person living at Casa 8, 486 Slate Bri Avenida, Heredia, San Jose 10101.

\textbf{Negative:} Leire Okafor is a person living at Casa 443, 601 Cypress Ursa ko'chasi, Namangan, Samarkand Region 140100, Uzbekistan.

\begin{center}
\small
\begin{tabular}{lccc}
\toprule
Stage & Pos. Similarity & Neg. Similarity & Gap \\
\midrule
Base Model & 0.8088 & 0.8346 & -0.0258 \\
Fine-Tuned & 0.9088 & 0.5801 & +0.3287 \\
\bottomrule
\end{tabular}
\end{center}

\paragraph{Example 5}

\textbf{Anchor:} Uche Eriksen is a person living at Apartamento 412, 430 Marigold Zuri Jalan, Yogyakarta, West Java 40111, Indonesia.

\textbf{Positive:} Uche Eriksen is a person living at 430 Marigold Zuri Jalan, Apartamento 412, Yogyakarta, West Java 40111, Indonesia.

\textbf{Negative:} Uche Eriksen is a person living at Apartamento 412, 430 Marigold Zuri Jalan, Different City, Different Region.

\begin{center}
\small
\begin{tabular}{lccc}
\toprule
Stage & Pos. Similarity & Neg. Similarity & Gap \\
\midrule
Base Model & 0.9915 & 0.9074 & +0.0841 \\
Fine-Tuned & 0.9905 & 0.5701 & +0.3363 \\
\bottomrule
\end{tabular}
\end{center}

% ============================================================
% BGE BUSINESS EXAMPLES
% ============================================================

\subsection{BAAI/bge-base-en-v1.5: Business Entities}

\paragraph{Example 1}

\textbf{Anchor:} Society of Talnori specialty milling OU is a firm located at Buroo 131, 398 Cobalt Tovi tanav, Parnu, Tartu County 51004, Estonia.

\textbf{Positive:} Socety of Talnori specialty milling OU is a firm located at Buroo 131, 398 Cobalt Tovi tnv, Parnu, Tartu County, Estonia.

\textbf{Negative:} Society of Talnori specialty milling OU is a firm located at Buroo 131, 398 Cobalt Tovi tanav, Al Ain, Central District 10035, United Arab Emirates.

\begin{center}
\small
\begin{tabular}{lccc}
\toprule
Stage & Pos. Similarity & Neg. Similarity & Gap \\
\midrule
Base Model & 0.9821 & 0.9241 & +0.0579 \\
Fine-Tuned & 0.9641 & -0.3536 & +1.3177 \\
\bottomrule
\end{tabular}
\end{center}

\paragraph{Example 2}

\textbf{Anchor:} Society of Xenovara bicycle components Ltd is a firm located at Suite 288, 334 Clover Sena Lane, Bristol, England YO1 7PR, United Kingdom.

\textbf{Positive:} Socety of Xenovara bicycle components Ltd is a firm located at Ste 288, 334 Clover Sena Ln, Bristol, England 7PR, United Kingdom.

\textbf{Negative:} Society of Xenovara bicycle components Ltd is a firm located at Suite 288, 334 Clover Sena Lane, Osaka, Central District 10013, Japan.

\begin{center}
\small
\begin{tabular}{lccc}
\toprule
Stage & Pos. Similarity & Neg. Similarity & Gap \\
\midrule
Base Model & 0.9678 & 0.9009 & +0.0669 \\
Fine-Tuned & 0.9576 & -0.3459 & +1.3034 \\
\bottomrule
\end{tabular}
\end{center}

\paragraph{Example 3}

\textbf{Anchor:} Society of Lunvara geotechnical drilling S.R.L. is a firm located at Oficina 168, 302 Elm Riv Avenida, Heredia, San Jose 10101, Costa Rica.

\textbf{Positive:} Socety of Lunvara geotechnical drilling SRL is a firm located at Of. 168, 302 Elm Riv Av., Heredia, San Jose, Costa Rica.

\textbf{Negative:} Society of Lunvara geotechnical drilling SRL is a firm located at Oficina 168, 302 Elm Riv Avenida, Bedok, Central District 10002, Singapore.

\begin{center}
\small
\begin{tabular}{lccc}
\toprule
Stage & Pos. Similarity & Neg. Similarity & Gap \\
\midrule
Base Model & 0.9536 & 0.8815 & +0.0721 \\
Fine-Tuned & 0.9605 & -0.3210 & +1.2815 \\
\bottomrule
\end{tabular}
\end{center}

\paragraph{Example 4}

\textbf{Anchor:} Society of Ivonori solar glazing sp. z o.o. is a firm located at Lokal 11, 366 Juniper Sor ulica, Lublin, Lesser Poland 30-001, Poland.

\textbf{Positive:} Socety of Ivonori solar glazing sp z oo is a firm located at Lokal 11, 366 Juniper Sor ul., Lublin, Lesser Poland 30-, Poland.

\textbf{Negative:} Society of Ivonori solar glazing sp z oo is a firm located at Lokal 11, 366 Juniper Sor ulica, Portland, Central District 10024, United States.

\begin{center}
\small
\begin{tabular}{lccc}
\toprule
Stage & Pos. Similarity & Neg. Similarity & Gap \\
\midrule
Base Model & 0.9661 & 0.8799 & +0.0862 \\
Fine-Tuned & 0.9506 & -0.3409 & +1.2916 \\
\bottomrule
\end{tabular}
\end{center}

\paragraph{Example 5}

\textbf{Anchor:} Society of Perrnori workwear manufacturing MChJ is a firm located at Ofis 371, 462 Papyrus Ursa ko'chasi, Bukhara, Samarkand Region 140100, Uzbekistan.

\textbf{Positive:} Socety of Perrnori workwear manufacturing MChJ is a firm located at Of. 371, 462 Papyrus Ursa ko'chasi, Bukhara, Samarkand Region, Uzbekistan.

\textbf{Negative:} Society of Perrnori workwear manufacturing MChJ is a firm located at Ofis 371, 462 Papyrus Ursa ko'chasi, Munich, Central District 10057, Germany.

\begin{center}
\small
\begin{tabular}{lccc}
\toprule
Stage & Pos. Similarity & Neg. Similarity & Gap \\
\midrule
Base Model & 0.9854 & 0.9246 & +0.0608 \\
Fine-Tuned & 0.9862 & -0.2525 & +1.2387 \\
\bottomrule
\end{tabular}
\end{center}

% ============================================================
% BGE PERSON EXAMPLES
% ============================================================

\subsection{BAAI/bge-base-en-v1.5: Person Entities}

\paragraph{Example 1}

\textbf{Anchor:} Gita Gyamfi is a person living at Lot 241E, 629 Cedar Vela straat, Breda, Gelderland 6811 AA, Netherlands.

\textbf{Positive:} Gita Gyamfi is a person living at 629 Cedar Vela straat, Lot 241E, Breda, Gelderland 6811 AA, Netherlands.

\textbf{Negative:} Gita Gyamfi is a person living at Lot 241E, 629 Cedar Vela straat, Different City, Different Region.

\begin{center}
\small
\begin{tabular}{lccc}
\toprule
Stage & Pos. Similarity & Neg. Similarity & Gap \\
\midrule
Base Model & 0.9931 & 0.9250 & +0.0680 \\
Fine-Tuned & 0.9613 & 0.6284 & +0.3328 \\
\bottomrule
\end{tabular}
\end{center}

\paragraph{Example 2}

\textbf{Anchor:} Zoran Deshmukh is a person living at Apartamento 238, 384 Seabird Mav Gudamj, Choibalsan, Orkhon 61020, Mongolia.

\textbf{Positive:} Zoran Deshmukh is a person living at 384 Seabird Mav Gudamj, Apartamento 238, Choibalsan, Orkhon 61020, Mongolia.

\textbf{Negative:} Zoran Deshmukh is a person living at Apartamento 238, 384 Seabird Mav Gudamj, Different City, Different Region.

\begin{center}
\small
\begin{tabular}{lccc}
\toprule
Stage & Pos. Similarity & Neg. Similarity & Gap \\
\midrule
Base Model & 0.9919 & 0.9141 & +0.0778 \\
Fine-Tuned & 0.9316 & 0.5891 & +0.3425 \\
\bottomrule
\end{tabular}
\end{center}

\paragraph{Example 3}

\textbf{Anchor:} Gita Gyamfi is a person living at Lot 241E, 629 Cedar Vela straat, Breda, Gelderland 6811 AA, Netherlands.

\textbf{Positive:} Gita Gyamfi is a person living at Lt 241E, 629 Cedar Vela straat, Breda, Gelderland 6811 AA, Netherlands.

\textbf{Negative:} Gita Gyamfi is a person living at Unit 999, 123 Fake Street, Breda, Gelderland 6811 AA, Netherlands.

\begin{center}
\small
\begin{tabular}{lccc}
\toprule
Stage & Pos. Similarity & Neg. Similarity & Gap \\
\midrule
Base Model & 0.9852 & 0.9056 & +0.0796 \\
Fine-Tuned & 0.9937 & 0.6649 & +0.3288 \\
\bottomrule
\end{tabular}
\end{center}

\paragraph{Example 4}

\textbf{Anchor:} Tenzin Etxeberria is a person living at Apartamento 480, 315 Sorrel Gra rue, Liege, Wallonia 5000, Belgium.

\textbf{Positive:} Tenzin Etxeberria is a person living at Apartamento 480, 315 Sorrel Gra rue, Liege, Wallonia 5000, Belgium (exact match).

\textbf{Negative:} Tenzin Etxeberria is a person living at Apartamento 480, 315 Sorrel Gra rue, Parana 80010-110, Brazil.

\begin{center}
\small
\begin{tabular}{lccc}
\toprule
Stage & Pos. Similarity & Neg. Similarity & Gap \\
\midrule
Base Model & 1.0000 & 0.9198 & +0.0802 \\
Fine-Tuned & 1.0000 & 0.6729 & +0.3271 \\
\bottomrule
\end{tabular}
\end{center}

\paragraph{Example 5}

\textbf{Anchor:} Mirela Haddad is a person living at Lot 67C, 583 Anise Ivo rue, Nimes, Occitanie 31000, France.

\textbf{Positive:} Mirela Haddad is a person living at 583 Anise Ivo rue, Lot 67C, Nimes, Occitanie 31000, France.

\textbf{Negative:} Mirela Haddad is a person living at Lot 67C, 583 Anise Ivo rue, Different City, Different Region.

\begin{center}
\small
\begin{tabular}{lccc}
\toprule
Stage & Pos. Similarity & Neg. Similarity & Gap \\
\midrule
Base Model & 0.9881 & 0.9191 & +0.0690 \\
Fine-Tuned & 0.9379 & 0.6222 & +0.3157 \\
\bottomrule
\end{tabular}
\end{center}

% ============================================================
% SUMMARY WITHIN APPENDIX B
% ============================================================

\subsection{Average Discrimination Gap}

Table~\ref{tab:gap-summary} summarizes the average discrimination gap across
the representative business and person examples. The improvement is calculated
as the difference between the fine-tuned and pretrained average gaps.

\begin{table}[t]
  \caption{Average discrimination gap across representative examples.}
  \label{tab:gap-summary}
  \centering
  \begin{tabular}{lccc}
    \toprule
    Model / Entity Type & Avg. Base Gap & Avg. FT Gap & Improvement \\
    \midrule
    all-MiniLM-L6-v2 / Business
        & 0.0349 & 0.3999 & +0.3649 \\
    all-MiniLM-L6-v2 / Person
        & 0.0072 & 0.3511 & +0.3439 \\
    BAAI/bge-base-en-v1.5 / Business
        & 0.0688 & 1.2866 & +1.2178 \\
    BAAI/bge-base-en-v1.5 / Person
        & 0.0749 & 0.3294 & +0.2545 \\
    \bottomrule
  \end{tabular}
\end{table}

\subsection{Observations}

The representative comparisons reveal several consistent patterns:

\begin{itemize}
    \item \textbf{Improved discrimination:}
    Fine-tuning widens the positive--negative gap for every model and
    entity-type group shown, indicating sharper discrimination between
    true matches and near-duplicate distractors.

    \item \textbf{Strongest effect for BGE business entities:}
    The largest improvement is observed for BAAI/bge-base-en-v1.5 on
    business entities. The average gap increases from approximately
    $0.069$ for the pretrained model to $1.287$ after fine-tuning.

    \item \textbf{Substantial improvement for MiniLM:}
    For all-MiniLM-L6-v2, the average gap increases from $0.0349$ to
    $0.3999$ for business entities and from $0.0072$ to $0.3511$
    for person entities.

    \item \textbf{Correction of ranking failures:}
    The pretrained models can assign a higher similarity to a negative
    example than to the corresponding positive example. In the representative
    MiniLM cases, negative base-model gaps are observed for business
    Example 4 and Example 5, and for person Examples 2, 3, and 4.
    Fine-tuning changes these gaps to positive values.

    \item \textbf{Sensitivity to location conflicts:}
    Several examples preserve highly similar or identical names while
    changing the business or residential location. Fine-tuning substantially
    reduces the similarity assigned to these non-matching records,
    demonstrating improved sensitivity to identity and location conflicts.
\end{itemize}

\end{document}